\documentclass[prl,twocolumn,showpacs,amsmath,letter]{revtex4}
\usepackage{graphicx}

\newcommand{\Journal}[4]{#1 \textbf{#2}, #3 (#4)}

\begin{document}

\title{Effect of Polarized Current on the Magnetic State of an Antiferromagnet}

\author{Sergei Urazhdin}
\author{Nicholas Anthony}
\affiliation{Department of Physics, West Virginia University,
Morgantown, WV 26506}

\pacs{73.40.-c, 75.50.Ee, 75.60.Jk, 75.70.Cn}

\begin{abstract}
We provide evidence for the effects of spin polarized current on a
nanofabricated antiferromagnet incorporated into a spin-valve
structure. Signatures of current-induced effects include bipolar
steps in differential resistance, current-induced changes of
exchange bias correlated with these steps, and deviations from the
statistics expected for thermally activated switching of spin
valves. We explain our observations by a combination of spin torque
exerted on the interfacial antiferromagnetic moments, and
electron-magnon scattering in antiferromagnet.
\end{abstract}

\maketitle

Polarized current flowing through magnetic heterostructures can
change the magnetic state of ferromagnets (F) due to the spin
transfer (ST) effect~\cite{cornellorig}. It may find applications in
magnetic memory devices, field sensors, and microwave generation.
Despite advances in tunnel junction-based ST devices, their wide
scale application is still deterred by the large power consumption.
The efficiency of spin transfer into a nanomagnet F is determined by
the threshold current for the onset of current-induced magnetic
precession~\cite{slonczewski96}
\begin{equation}\label{it}
I_t=e2\pi m^2\alpha/(\hbar Vg),
\end{equation}
at small external magnetic field $H$. Here, $e$ is the electron
charge, $m$ and $V$ are the magnetic moment and the volume of the
nanomagnet, $\alpha$ is the Gilbert damping parameter, and $g$ is a
unitless parameter characterizing the efficiency of ST. $I_t$ is
usually close to the current that reverses the magnetization of F.

According to Eq.~\ref{it}, $I_t$ can be significantly reduced if F
has a small magnetic moment $m$. In nanopatterned magnets, a small
$m$ compromises the magnetic stability of devices. However, more
complex magnetic systems such as antiferromagnets (AF) can combine
vanishing $m$ with a significant magnetic anisotropy, and may thus
provide both stability and low operating power. Additionally, the
possibility to change the magnetic structure of AF by current is
attractive for devices utilizing AF for pinning the magnetization of
the adjacent F, due to the exchange bias (EB) effect. ST in
antiferromagnets has been predicted~\cite{nunez}, but experimental
studies have been
inconclusive~\cite{ohno},~\cite{emley},~\cite{tsoiaf}. Such studies
may provide information about the role of electron-magnon scattering
in ST phenomena, the nature of enhanced magnetic damping in F/AF
bilayers~\cite{damping}, and the properties of the interfacial
magnetic moments central to EB.

\begin{figure}
\includegraphics[scale=0.47]{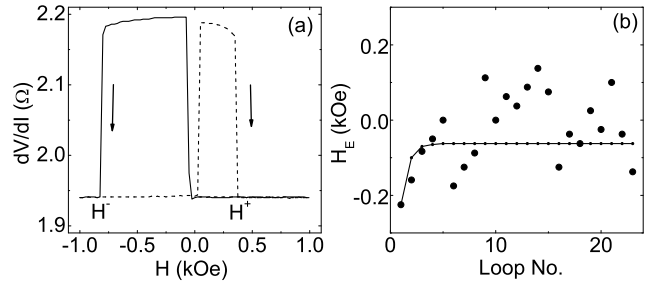}
\caption{\label{fig1} (a) Hysteresis loop of EB nanopillar acquired
at $4.5$~K immediately after cooldown at $H=3$~kOe from $350$~K.
Reversal fields $H^+$ and $H^-$ of Py(5) are labeled. (b) Exchange
bias $H_E$ {\it vs.} the hysteresis loop number in EB nanopillar
(symbols) and a 5$\times$5~mm sample with magnetic multilayer
structure identical to the EB nanopillar (curve).}
\end{figure}

We report on the effect of polarized current on nanoscale AF
elements incorporated in a giant magnetoresistive (MR) structure
F$_1$/N/F$_2$/AF, with the top F$_2$/AF bilayer patterned into an
120$\times$60 nm nanopillar. These samples are labeled EB
nanopillars. Several important behaviors of our samples can be
attributed to the effect of current on AF. First, EB an be changed
by applying a pulse of current. The changes are non-monotonic and
asymmetric with respect to the directions of current and applied
field, which eliminates Joule heating as their origin. Second, the
peaks in the current-induced EB are correlated with the onset of
magnetic dynamics, indicating its importance for defining the
magnetic state of AF. Finally, switching is inconsistent with the
standard thermal activation model, indicating a complex
current-induced magnetic state of AF.

We tested 8 samples with the structure
Py(30)Cu(10)Py(5)Fe$_{50}$Mn$_{50}$(t)Cu(1)Au(10), where
Py=Permalloy=Ni$_{80}$Fe$_{20}$, and $1.5\le t\le 4$. All
thicknesses are in nanometers. Samples with $t>2$ exhibited complex
reversal patterns and signatures of inhomogeneous magnetic states.
We interpret these behaviors in terms of the local variations of
exchange interaction at the F/AF interface. On the other hand,
samples with $t=1.5$ exhibited the important features of EB
associated with FeMn, while retaining single domain behaviors
similarly to the samples not containing an AF layer (standard
samples). The results reported below were verified for two EB
samples with $t=1.5$. Current $I>0$ flowed from the extended to the
patterned Py layer, and $H$ was along the nanopillar easy axis.

The MR of the EB sample shown in Fig.~\ref{fig1}(a) was
$0.26$~$\Omega$ at $T=4.5~K$, among the largest reported for
metallic
structures~\cite{cornellorig},~\cite{emley},~\cite{myprl},~\cite{cornellnature},~\cite{krivorotov}.
The data were acquired immediately after cooldown at $H=3$~kOe from
$350$~K. Field-induced reversals occurred in a single step at all
temperatures $T$ between $4.5$~K and room temperature RT=$295$~K.
Upward jumps of $dV/dI$ in Fig.~\ref{fig1}(a) at small $H=\pm30$~Oe
are caused by the reversals of the extended Py(30) layer from the
parallel (P) state of the Py layers into the antiparallel (AP)
state. The asymmetric downward jumps at large $H^+>0$ and $H^-<0$
(labeled in Fig.~\ref{fig1}(a)) are caused by the reversal of the
patterned Py(5) layer. The asymmetry is due to FeMn, and can be
characterized by the effective EB field $H_E=(H^++H^-)/2=-225$~Oe.
The coercivity is $H_C=(H^+-H^-)/2=588$~Oe.

Very thin FeMn layers usually become magnetically unstable due to
the reduced anisotropy energy~\cite{nogues}. Indeed, repeated
cycling of $H$ yielded fluctuating values of $H_E$ (symbols in
Fig.~\ref{fig1}(b)). This behavior was attributed to reorientation
of FeMn among a few metastable states. When similar measurements are
performed in an extended film, $H_E$ should quickly decay from the
initial value determined by the field cooling, due to the averaging
of magnetic fluctuations over a large number of AF grains. This
behavior was verified in a $5\times5$~mm$^2$ heterostructure
Py(5)Cu(10)Py(5)FeMn(1.5)Cu(1)Au(3) magnetically identical to the EB
nanopillars (curve in Fig.~\ref{fig1}(b)).

\begin{figure}
\includegraphics[scale=0.85]{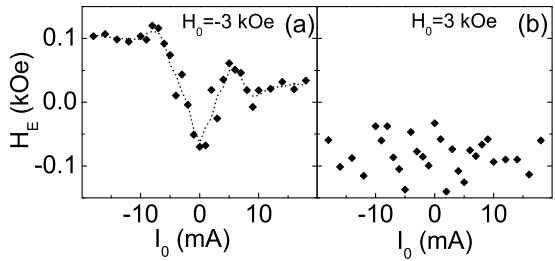}
\caption{\label{fig2} $H_E$ of EB sample measured at $I=0.3$~mA
after a 100~ms pulse of current $I_0$ applied at $H_0=-3$~kOe (a)
and $H_0=3$~kOe (b). Each point is an average of 10 measurements.
$H_0$ and/or $I_0$ were reversed after each group of measurements.
The dashed curve in (a) is guide to the eye.}
\end{figure}

The effect of current on the magnetic structure of AF can be
determined by applying a pulse of current $I_0$ at field $H_0$, and
subsequently measuring the hysteresis loop at a small $I$ not
affecting the magnetic state of the nanopillar. The results of such
measurements are shown in Fig.~\ref{fig2} for $H_0=-3$~kOe (a) and
$H_0=3$~kOe (b). These values of $H_0$ suppressed current-induced
reversal of the Py(5) nanopillar. The most prominent feature of the
data is a sharp increase of $H_E$ up to a peak at $I_0=-7.5$~mA, and
a smaller peak at $I_0=5$~mA in Fig.~\ref{fig2}(a), for
$H_0=-3$~kOe. The data in Fig.~\ref{fig2}(b) for $H_0=3$~kOe are
scattered, and do not exhibit a clear dependence on $I_0$.  If these
changes of $H_E$ were caused by Joule heating of FeMn, the data in
Figs.~\ref{fig2}(a) and (b) would mirror each other, at least for
large $I_0$. Additionally, the effect of heating would increase with
$I_0$, so a monotonic dependence instead of peaks would be expected.
Therefore, current must have a direct effect on the magnetic state
of FeMn.

\begin{figure}
\includegraphics[scale=0.81]{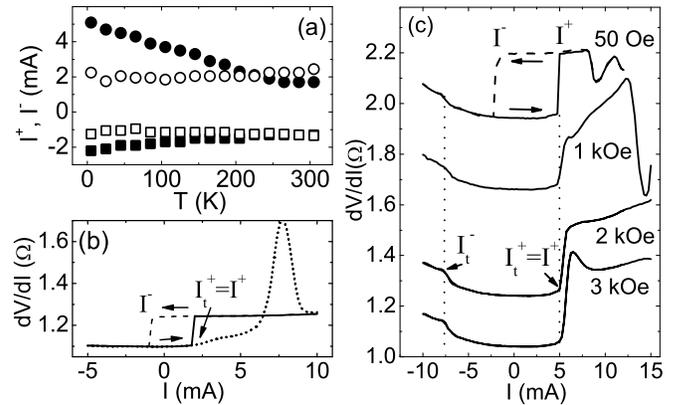}
\caption{\label{fig3} (a) Switching currents from P to AP state
(circles) and from P to AP state (squares) {\it vs.} $T$ for EB
nanopillar (solid symbols) and standard sample (open symbols), at
$H=50$~Oe. (b) $dV/dI$ {\it vs.} $I$ at $4.5$~K for standard sample,
with $H=50$ Oe (dashed and solid curves), and $363$~Oe (dotted
curve). (c) Same as (b), but for the EB sample at the labeled values
of $H$. Curves for $H>50$~Oe are offset for clarity.}
\end{figure}

Additional evidence for the effects on FeMn distinct from Joule
heating was provided by the measurements of current-induced
behaviors. At small $H$, Py(5) reversed into the P state at a
current $I^-$, and into the AP state at a current $I^+$, consistent
with the ST mechanism (top curve in Fig.~\ref{fig3}(c)). However,
the dependencies of $I^-$ and $I^+$ on $T$ reflected a strong
influence of FeMn on the current-induced reversals.
Fig.~\ref{fig3}(a) shows that $I^-$ and $I^+$ in the standard sample
did not significantly depend on $T$ (open symbols), but they
increased linearly with decreasing $T$ in the EB
nanopillars~\cite{shunting}. This result is consistent with the
previously seen enhancement of magnetic damping due to
AF~\cite{emley},~\cite{damping}.

Current-induced behaviors of the EB samples were different from the
standard ones at large $H$, as shown in Figs.~\ref{fig3}(b),(c). The
standard sample exhibited two previously established
features~\cite{myprl}. First, switching became reversible and
exhibited telegraph noise, resulting in a peak in $dV/dI$ at
$I>I^+_t$ (dotted curve in Fig.~\ref{fig3}(b)). The peak rapidly
shifted to higher $I$ at larger $H$. The second feature was a weak
increase of $dV/dI$ at $I>I^+_t\approx I^+$. This feature is
associated with the current-induced precession of the Py(5)
magnetization~\cite{cornellnature}.

The EB nanopillars did not exhibit a reversible switching peak at
any $H$. The $H=1$~kOe data in Fig.~\ref{fig3}(c) show a step at
$I^+_t$, and an approximately linear increase at $I>I^+_t$ until a
drop at $12$~mA. At $2$~kOe and $3$~kOe, the step at $I^+_t$ is
larger, and the increase of $dV/dI$ at higher $I$ is smaller.
Because the position of the step does not significantly depend on
$H$, it can be attributed to the current-induced precession of
Py(5), as confirmed by the time-resolved measurements described
below. The step at $I^+_t$ is correlated with a peak of $H_E$ in
Fig.~\ref{fig2}(a). Similarly, a smaller step in $dV/dI$ at
$I=-7.5$~mA, labeled $I^-_t$ in Fig.~\ref{fig3}(c), is also
correlated with a peak of $H_E$ in Fig.~\ref{fig2}(a) at $I<0$. Both
steps are attributed to the current-induced magnetic dynamics, which
must therefore play an important role in current-induced EB. To
confirm that these steps are associated with the FeMn layer, $T$ was
increased to RT. Both steps disappeared, and the reversible
switching peak was observed.

\begin{figure}
\includegraphics[scale=0.8]{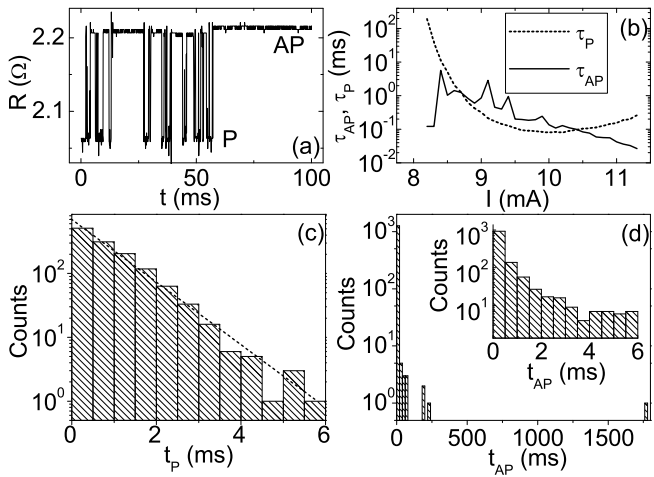}
\caption{\label{fig4} (a) $R$ {\it vs} $t$ for EB nanopillar at
$I=9$~mA, $H=900$~Oe. (b) Average dwell times in the P state
$\tau_P$ (dashed curve) and the AP state $\tau_{AP}$ (solid curve)
{\it vs.} $I$, at $H=900$~Oe. (c) Distribution of dwell times in the
P state at $I=9$~mA and $H=900$~Oe. Dashed line is exponential fit
with characteristic time $\tau_0=1.1$~ms. (d) Same as (c), for the
AP state. Inset shows the distribution for the shorter time scale.}
\end{figure}

To identify the origin of the high-field current-induced behaviors
in Fig.~\ref{fig3}(c), we performed time-resolved measurements of
resistance $R$. Fig.~\ref{fig4}(a) shows an example of a time trace
acquired at $I=9$~mA and $H=900$~Oe. The data show telegraph noise
switching between P and AP states, qualitatively similar to the
standard sample. However, there are significant quantitative
differences. First, the EB nanopillar occasionally spent a
significant continuous period of time in the AP state, which was not
observed in the standard samples. Second, the average dwell times in
the P state ($\tau_{P}$) and the AP state ($\tau_{AP}$) exhibited a
nonmonotonic dependence on $I$, as illustrated in
Fig.~\ref{fig4}(b)~\cite{tauap}. In contrast, the standard sample
exhibited a monotonic increase of $\tau_{AP}$ and a decrease of
$\tau_{P}$ with increasing $I$. The data for the standard sample
were consistent with the Neel-Brown model of thermal magnetic
activation~\cite{zhang}.

The dependencies of average dwell times on $I$ account for the
approximately linear increase of $dV/dI$ at $I>I^+_t$ in the EB
samples. The origin of these dependencies can be better understood
by plotting the distributions of dwell times in the P state ($t_P$)
and the AP state ($t_{AP}$) (Figs.~\ref{fig4}(c),(d)). $t_P$ closely
followed an exponential distribution with a decay time
$\tau_0=1.1$~ms at $I=9$~mA and $H=900$~Oe. Similar distributions
were obtained for both $t_P$ and $t_{AP}$ in the standard samples,
consistently with the thermal activation model~\cite{zhang}.
However, the distribution of $t_{AP}$ for the EB nanopillar was
clearly not exponential due to occasionally long dwell times in the
AP state. Of the total $10$~sec time interval of data analyzed in
Fig.~\ref{fig4}(d), the system spent one continuous $1.8$~s long
interval, and three $200$~ms long intervals in the AP state. The
remaining dwell times were predominantly less than $1$~ms, as shown
in the inset. As $I$ was increased, the intervals of long $t_{AP}$
gradually disappeared, resulting in decreasing $\tau_{AP}$ in
Fig.~\ref{fig4}(b). Simple enhancement of magnetic damping and/or
the anisotropy induced by FeMn cannot account for the deviations
from the standard exponential distributions. Therefore, this
behavior is the strongest evidence in our data for the effect of
$I>0$ on the magnetic state of FeMn, which results in occasional
enhancement of Py(5) stability in the AP state.

Before discussing the mechanisms underlying the observed effects of
current on the AF/F bilayer, we summarize the current-induced
behaviors distinguishing the EB samples from the standard ones: i)
The switching currents depend linearly on $T$, ii) The effect of
current on $H_E$ is asymmetric with respect to the directions of $H$
and $I$, iii) Current-induced $H_E$ exhibits peaks correlated with
the steps in $dV/dI$, iv) Reversible switching exhibits occasionally
long dwell times in the AP state.

We analyze our data using a model accounting for the contributions
of both the spin current and the spin accumulation to ST at magnetic
interfaces~\cite{fert}. The parameter $g$ in Eq.~\ref{it}
characterizing ST at the interface N/F$_2$ is expressed through the
known material parameters
\begin{equation}\label{st}
g=(j_{F2}+v_{F2}m_{F2}/8)e/j,
\end{equation}
where $j$ is the current density, $v_{F2}\approx0.3\times10^6$~m/s
is the average Fermi velocity in F$_2$, $j_{F2}$ and $m_{F2}$ are
the spin current density and spin accumulation density,
respectively, just inside F$_2$ at the N/F$_2$ interface. Spin
flipping in the Cu spacer is neglected in Eq.~\ref{st}. The spin
accumulation and spin current density were calculated
self-consistently throughout the multilayer using the Valet-Fert
model~\cite{valetfert}. Material parameters known from MR
measurements and electron photoemission were
used~\cite{photoemission},~\cite{gmr}. Eq.~\ref{st} yielded $g=0.89$
for the P state, and $g=-1.88$ for the AP state. Inserting those
values into Eq.~\ref{it}, we obtained good agreement with the
switching currents at $4.5$~K by assuming $\alpha=0.12$. This value
is in general agreement with $\alpha=0.03$ for the standard
samples~\cite{krivorotov}, and the enhancement of damping due to
FeMn, as determined from the temperature dependence of switching
currents (Fig.~\ref{fig3}(a)).

A calculation for the Py/FeMn interface yielded a positive $g=1.2$
for the P state, implying that FeMn moments that are noncollinear to
Py experience a spin torque favoring their parallel configuration
with Py for $I<0$, and antiparallel configuration for $I>0$.
Interfacial Mn moments tend to align antiparallel to the adjacent F,
while Fe moments align parallel to it~\cite{femn}. The latter have a
larger magnetic anisotropy and likely dominate the EB. The following
picture of ST at the Py/FeMn interface then emerges. The positive
value of $g$ at the Py/FeMn interface implies that ST acting on the
interfacial Fe moments enhances EB for $I_0<0$ and suppresses it for
$I_0>0$, explaining the asymmetry with respect to the current
direction in Fig.~\ref{fig2}(a), for $H_0=-3$~kOe. A step in $dV/dI$
at $I^-_t$ is likely caused by the current-induced precession of the
stable Fe moments that remain antiparallel to the Py magnetization,
consequently inducing dynamical Py response resulting in MR.

The asymmetric current-induced behaviors in Fig.~\ref{fig2}(a) are
superimposed on the symmetric enhancement of EB, independent of the
current direction. We propose two possible mechanisms. First,
electron-magnon scattering at the Py/FeMn interface may activate
transitions of the FeMn magnetic moments into the stable orientation
dictated by the magnetization of the adjacent Py. A second, probably
weaker, effect may be due to a torque on the Fe moments exerted by
the Oersted field of $I$, assisting their rotation into the
direction parallel to the Py moments. The peaks at $I^+_t$ and
$I^-_t$ support the activation picture, since magnetic dynamics
excited in FeMn either directly by ST, or indirectly through
interaction with precessing Py, should assist in activating the
transition of the AF magnetic moments into a stable configuration.

To interpret the lack of significant current dependence of EB on
$I_0$ for the $H_0=3$~kOe data, we note that the average $H_E$ in
Fig.~\ref{fig2}(b) is independent of $I_0$, and is similar but
opposite in sign to the largest values obtained for $H_0=-3$~kOe.
Therefore, at $H_0=3$~kOe the system simply reverts to a stable
configuration defined by the in-field cooldown. This state with
higher anisotropy is not significantly affected by $I$.

We also propose an interpretation for the non-exponential
distribution of $t_{AP}$ in Fig.~\ref{fig4}(d).  A simple
enhancement of the AP state stability due to EB cannot be
responsible, because then $t_P$ would also exhibit a similar effect.
Therefore, the occasionally long $t_{AP}$ are likely caused by the
current-induced enhancement of magnetic stability, efficient only in
the AP state at $I>0$, but not in the P state. Since $I>0$
suppresses the fluctuations of the magnetic layer in the AP state,
due to the ST in the standard samples~\cite{zhang}, a similar
combined effect of current on both F and AF layers likely takes
place in the EB samples. It is efficient only for some of the
magnetic configurations that FeMn acquires due to the fluctuations,
specifically those minimizing the current-induced precession of Fe
moments, resulting only in occasional enhancement of $t_{AP}$.

In summary, we demonstrated that spin polarized electron current
affects the magnetic state of an antiferromagnet. The effect is
distinct from Joule heating, and is explained by a combination of a
direct current-induced excitation of antiferromagnetic moments by
spin transfer, and electron-magnon scattering at the magnetic
interface. This mechanism may be useful for establishing exchange
bias in nanoscale magnetoelectronic devices.

We acknowledge discussions with D. Lederman, K.-J. Lee, I. Moraru,
N.O. Birge, and J. Bass.


\begin{thebibliography}{99}
\bibitem{shunting} The slight increase of $I^-$, $I^-$
at $T>250K$ is caused by the shunting currents through the Si
substrate.
\bibitem{tauap} fluctuations of $\tau_{AP}$ in
Fig.~\ref{fig4}(b) are caused by occasionally large $t_{AP}$
discussed in the text.
\end{thebibliography}
\end{document}